\documentclass[aps,pre,twocolumn,superscriptaddress]{revtex4}
\usepackage{graphicx}
\usepackage{amssymb}
\usepackage{amsmath}
\usepackage{url}
\begin{document}
\title{Specific heat anomalies of open quantum
systems}
\author{Gert-Ludwig Ingold}
\email{gert.ingold@physik.uni-augsburg.de}
\affiliation{Institut f\"ur Physik, Universit\"at Augsburg, D-86135 Augsburg}
\affiliation{Laboratoire Kastler Brossel, CNRS, ENS, UPMC, Campus Jussieu Case 74,
F-75252 Paris Cedex 05, France}
\author{Peter H\"anggi}
\author{Peter Talkner}
\affiliation{Institut f\"ur Physik, Universit\"at Augsburg, D-86135 Augsburg}

\begin{abstract}
The evaluation of the specific heat of an open, damped quantum system is a subtle
issue. One possible route is based on the thermodynamic partition function which is
the ratio of the partition functions of system plus bath and of the bath alone.
For the free damped particle it has been shown, however, that the ensuing
specific heat may become negative for appropriately chosen
environments. Being an open system this quantity then naturally must
be interpreted as the change of the specific heat obtained as the
difference between the specific heat of the heat bath coupled 
to the system degrees of freedom and the 
specific heat of the bath alone. While this difference may become negative,
the involved specific
heats themselves are always positive; thus, the known thermodynamic
stability criteria are perfectly guaranteed. For a damped quantum
harmonic oscillator, instead of negative values, under appropriate
conditions one can observe a dip in the difference of specific heats
as a function of temperature. Stylized minimal models containing a
single oscillator heat bath are employed to elucidate the occurrence
of the anomalous  temperature dependence of the corresponding
specific heat values. Moreover, we comment on the consequences for the
interpretation of the density of states based on the thermal partition
function.
\end{abstract}


\maketitle

\section{Introduction}

An open classical system in contact with a heat bath can often be
modeled in terms of a Langevin dynamics with constant friction and
white Gaussian noise sources obeying a fluctuation-dissipation theorem
\cite{Coffey,Zwanzig,RMP90}. A remarkable feature  then
is the circumstance that the equilibrium statistics of the open
classical system 
turns out to be {\it independent} of the coupling
strength between the system and the heat bath. 
In other words, the canonical
equilibrium for a classical damped Langevin dynamics agrees with 
the canonical  equilibrium of the isolated system. 
This feature is rooted in the fact
that in this case the so termed ``Hamiltonian of mean force'' is still
given by the bare system Hamiltonian \cite{Campisi}. 
In clear contrast, this property in general no longer holds true for open  
systems in the quantum regime beyond the weak-coupling limit
\cite{hangg05}. In particular, the canonical equilibrium
state of an open quantum system then typically involves an explicit
dependence on the system-bath coupling strength. 

Motivated by this fact, the study of
the specific heat beyond the weak-coupling limit has recently received
considerable attention, in particular in view of the validity of the Third Law
of thermodynamics \cite{hangg06,hoerh07,bandy08,hangg08,wang08,kumar09,bandy09}. Apart from
fundamental thermodynamical questions the study of the specific heat in the
quantum regime is also of interest because it can be related to entanglement
properties \cite{wiesn08}.

Recently, two different routes towards the evaluation of a specific heat were proposed and
discussed \cite{hangg06,hangg08}. One possibility is based on the thermal expectation
value of the Hamiltonian describing the  isolated system. Another approach, on which
we will focus in this paper, starts out from the thermodynamic partition function of the
dissipative system  \cite{hangg06,hangg08,feynm63,feynm72,calde83,grabe84,legge87,grabe88,ford88,hanke95,qtad98,ingol02,ford07}
\begin{equation}
\label{eq:partitionfunction}
 Z = \frac{\mathrm{Tr}_\mathrm{S+B}[\exp(-\beta H)]}
     {\mathrm{Tr}_\mathrm{B}[\exp(-\beta H_\mathrm{B})]}
\end{equation}
where the total Hamiltonian
\begin{equation}
 H = H_\mathrm{S}+H_\mathrm{B}+H_\mathrm{SB}
\end{equation}
consists of terms describing the system, the bath, and the system-bath
coupling, respectively. In the absence of a coupling between system and bath,
$Z$ reduces to the partition function of the system. The partition function
(\ref{eq:partitionfunction}) appears naturally in the Feynman-Vernon
approach to dissipative systems \cite{feynm72,calde83,legge87,grabe88} and
can be related to equilibrium properties of the system \cite{grabe84,ingol02}.

From (\ref{eq:partitionfunction}), one obtains by means of standard thermodynamic
relations a specific heat \cite{footnote1}, reading
\begin{equation}
\label{eq:specificheat}
 C = k_\mathrm{B}\beta^2\frac{\partial^2}{\partial\beta^2}\ln(Z)\,.
\end{equation}
Here, $k_\mathrm{B}$ is the Boltzmann constant and the temperature $T$ appears
through $\beta=1/k_\mathrm{B}T$.

\begin{figure}
\begin{center}
\includegraphics[width=\columnwidth]{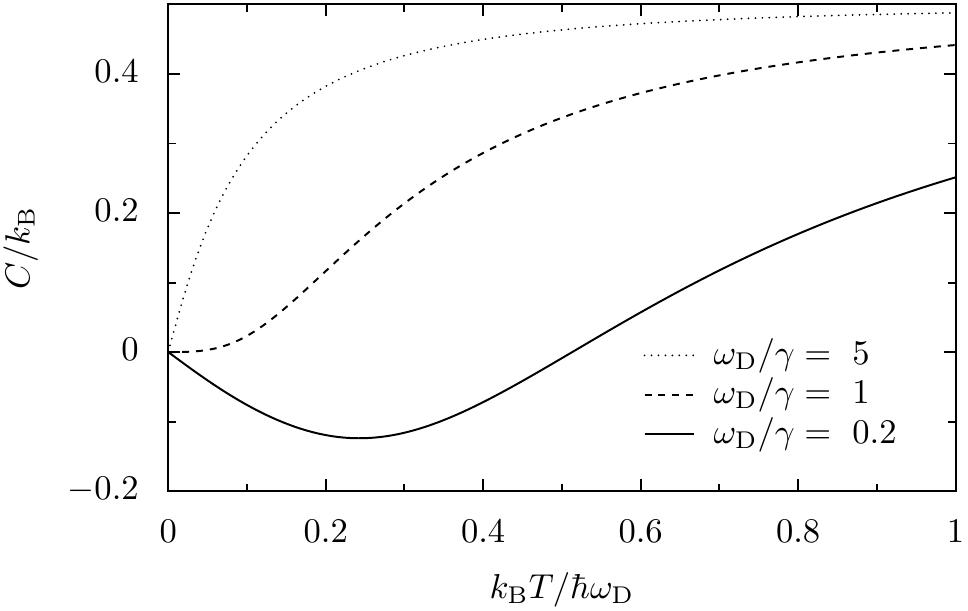}
\end{center}
\caption{The specific heat as defined in (\ref{eq:specificheat})
  for a free damped particle with a Drude damping kernel (\ref{eq:drude}) is
shown as a function of temperature for $\omega_\mathrm{D}/\gamma=0.2,1,$ and 5.
Note that in contrast to Ref.~\cite{hangg08}, temperature is given in units
of the cutoff frequency $\omega_\mathrm{D}$ instead of the damping
strength $\gamma$.}
\label{fig:c}
\end{figure}

In the following, we will assume the bath to consist of harmonic oscillators
and the coupling to be bilinear in system and bath coordinates
\cite{hangg05}. In such a framework Ref.~\cite{hangg08} found
for the damped free particle, that under certain circumstances the specific heat
(\ref{eq:specificheat}) can become negative. In the case of the Drude model,
for example, where the Laplace transform of the damping kernel is given by
\begin{equation}
\label{eq:drude}
\hat\gamma(z)= \frac{\gamma\omega_\mathrm{D}}{z+\omega_\mathrm{D}}\,,
\end{equation}
the specific heat exhibits negative values at low temperatures if the
damping constant $\gamma$ exceeds the cut-off frequency $\omega_\mathrm{D}$,
i.e. $\gamma>\omega_\mathrm{D}$.
This behavior is depicted in Fig.~\ref{fig:c}.
For a general damping kernel, a negative specific heat will appear at low
temperatures if $\hat\gamma'(0)<-1$, see Ref.~\cite{hangg08}. 
No negative specific heat was found if the
free particle is replaced by a harmonic oscillator.

In this paper, we will elucidate the origin of the negative specific heat for
the case of the free damped particle and its absence for the damped harmonic
oscillator by considering two minimal models. Before doing so, a general
comment on the appearance of a negative specific heat is appropriate.

The rationale behind the definition of a specific heat based on the thermodynamic partition
function (\ref{eq:partitionfunction}) is that this partition function should be
associated with an effective description of an open system.  From this point
of view, a negative specific heat may appear disturbing because it raises doubts as to
the thermodynamic stability of the system.

However, the meaning of the thermodynamic partition function (\ref{eq:partitionfunction}) of the open system can
be better understood from the point of view of its finite coupling to the heat bath. The specific
heat (\ref{eq:specificheat}) can in fact be expressed as
\begin{equation}
C = C_\mathrm{S+B}-C_\mathrm{B}\,,
\label{C_diff}
\end{equation}
where $C_\mathrm{S+B}$ is the specific heat of system and heat bath while $C_\mathrm{B}$
is the specific heat of the heat bath alone.

Therefore, $C$ describes the change
of the specific heat when the heat bath is enlarged by coupling it to system degrees of
freedom. This difference expression in (\ref{C_diff}) must 
be expected on physical grounds when dealing with an {\it
  open} system that is not heat-isolated from its environment. For
example, take a certain amount of an agent within a container: The
established experimental  procedure to determine the specific heat of
this agent is first to measure the specific heat of the empty
container and to subtract this value from the measured specific heat
of the combined system to finally arrive at the specific heat of the
agent alone. For a macroscopic amount of the agent this value is truly
agent specific, i.e. independent of the particular nature of the container and its
interaction with the agent (provided the interaction of the agent with
the container is short ranged). For a nanoscopic system, however, the
energy that is contained in the system-bath coupling typically cannot
be neglected and consequently will influence the specific heat of such
a system.     
Notably the thermodynamic internal energy of
the open system  also typically differs from the thermal expectation
of the bare system Hamiltonian \cite{Campisi,hoerh07}. As a
consequence, the  values of the specific heat evaluated along such
different routes then differ as well \cite{hangg06,hangg08}. 

The coupling can thus result in a negative specific heat of the
open quantum system. However, the involved specific heats, i.e. those
of system plus bath on the one hand and of the bath alone on the other
hand, are each positive so that no issues concerning the
thermodynamic stability arise \cite{Callen}. Such physical situations
are not uncommon as evidenced by recent discussions in the context of
the Casimir effect \cite{hoye03}, the multichannel Kondo effect
\cite{flore04}, or the physics of mesoscopic superconductors that
contain magnetic impurities \cite{prusc08}.

\section{Free particle}
\label{sec:free}
\begin{figure}
\begin{center}
\includegraphics[width=\columnwidth]{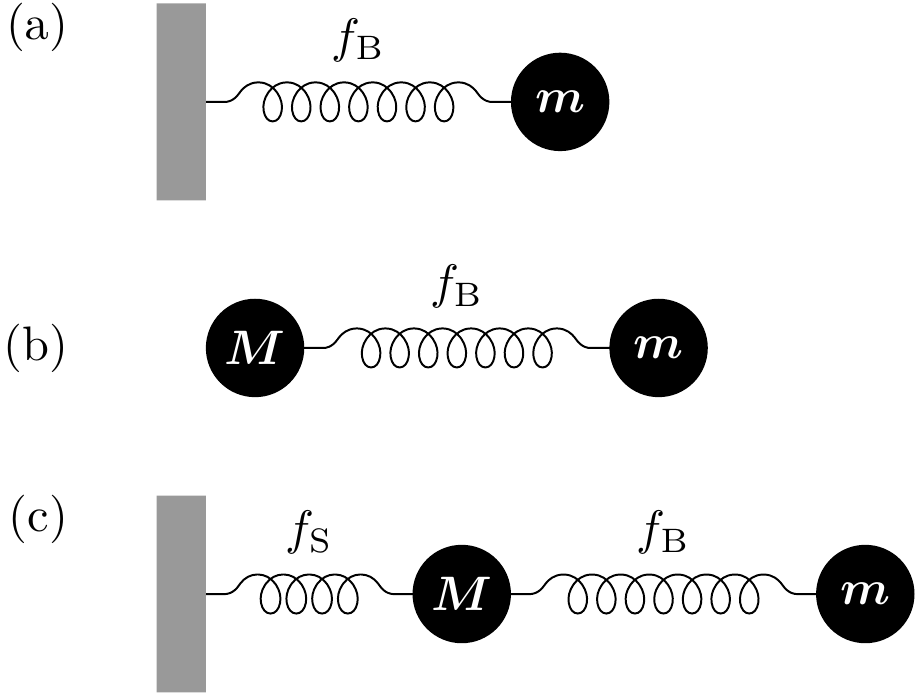}
\end{center}
\caption{(a) One single bath oscillator represented by a mass $m$ harmonically coupled
to a wall of infinite mass. (b) Bath oscillator coupled to a free system
degree of freedom. (c) Bath oscillator coupled to a harmonically bound system degree
of freedom. The spring constants of the bath and system oscillator are
denoted by $f_\text{B}$ and $f_\text{S}$, respectively.}
\label{fig:toymodels}
\end{figure}

In order to elucidate the appearance of a negative specific heat (\ref{eq:specificheat})
it is sufficient to consider a stylized, minimal model where the
``bath'' consists of only a single 
degree of freedom described by the Hamiltonian
\begin{equation}
H_\mathrm{B} = \frac{p^2}{2m}+\frac{f_\text{B}}{2}q^2
\label{HB}
\end{equation}
where $f_\text{B}$ denotes the spring constant.
In the following, we will study a system governed by the Hamiltonian
\begin{equation}
H_\mathrm{S} = \frac{P^2}{2M}+\frac{f_\text{S}}{2}Q^2\, ,
\end{equation}
both in the cases of a free particle (spring constant $f_\text{S}=0$) and of a
harmonic oscillator ($f_\text{S}>0$). The coupling Hamiltonian is given by
\begin{equation}
H_\mathrm{SB} = -f_\text{B}qQ+\frac{f_\text{B}}{2}Q^2\, ,
\label{SB}
\end{equation}
where the last term renormalizes the potential in order to ensure translational
invariance in the case of the free particle. Figure~\ref{fig:toymodels} illustrates
our two minimal models. The starting point is a single bath oscillator
with mass $m$ as depicted
in Fig.~\ref{fig:toymodels}a. Coupling the system mass $M$ to the bath oscillator
leads to the harmonically coupled system of two masses shown in
Fig.~\ref{fig:toymodels}b.
If the system degree of freedom corresponds to a harmonic oscillator, we obtain
the mechanical system of Fig.~\ref{fig:toymodels}c.

It is quite obvious that an environment consisting of one degree of freedom does 
not suffice to replace any realistic heat bath. In particular, it does not lead 
to a truly dissipative behavior of the open system to which it couples. Nevertheless, 
it turns out that even such minimal bath models give rise to the same thermodynamic 
anomalies that are also encountered with more realistic, large environments.  

We begin our analysis with a free particle in contact with the
single-degree-of-freedom environment described by eqs. (\ref{HB}) and
(\ref{SB}). System and bath are assumed to stay in thermal equilibrium
with each other at the inverse temperature $\beta$. Hence, the density
matrix of the total system is given by a Gibbs state reading
\begin{equation}
\rho_{\text{SB}} = Z^{-1}_{\text{SB}} \exp \left
  [-\beta(H_{\text{S}}+ H_{\text{B}}+ H_{\text{SB}})  \right]
\label{rSB}
\end{equation}
where $Z_{\text{SB}} = \text{Tr} \exp \left
  [-\beta(H_{\text{S}}+ H_{\text{B}}+ H_{\text{SB}})  \right]$ denotes
the partition function of the total system.    

The partition function $Z_\mathrm{B} =\text{Tr} \exp\left
  [-\beta H_{\text{B}}  \right] $ of the isolated 
bath degree of freedom is given by
\begin{equation}
\label{eq:zb0}
Z_\mathrm{B} = \frac{1}{2\sinh\left(\displaystyle\frac{\hbar\beta\omega}{2}\right)}
\end{equation}
where
\begin{equation}
\label{omega}
\omega=\left(\frac{f_\text{B}}{m}\right)^{1/2}
\end{equation}
is the frequency of the bath oscillator. From (\ref{eq:specificheat}) the specific
heat of this bath follows as
\begin{equation}
\label{CB}
C_{B} = k_{B} \:g\!\left(\frac{\hbar \beta \omega}{2}\right)
\end{equation}
with the abbreviation
\begin{equation}
\label{eq:gabbrev}
g(x)=\left(\frac{x}{\sinh(x)}\right)^2.
\end{equation}
If we add the system degree of freedom in order to obtain the mechanical system shown
in Fig.~\ref{fig:toymodels}b, the partition function contains contributions of two
degrees of freedom related to the center-of-mass and the relative motion. The first one
is described by a free particle with an effective mass $m+M$ while the second degree of
freedom corresponds to a harmonic oscillator with effective mass $mM/(m+M)$ and the
frequency
\begin{equation}
\label{eq:omegabar}
\bar\omega = \left(1+\frac{m}{M}\right)^{1/2}\omega\,.
\end{equation}
As discussed in the introduction, negative values of the specific heat
occur for the Drude model if the damping strength exceeds the cut-off
frequency of the heat bath.  Within our minimal model the specific
heat may become negative 
if the mass ratio $m/M$ exceeds a value slightly above 4.
Then, $\bar\omega$ is significantly larger than $\omega$, a fact which will be
relevant for the discussion of the specific heat (\ref{eq:cfree}) below.

In order to obtain a well-defined partition function for the free particle, we
restrict its motion to a region of length $L$. This length is supposed to be
sufficiently large such that the energy level spacing can be neglected if
compared with the thermal energy $k_BT$ \cite{hangg08}. Under
this condition $L$ will turn out to be irrelevant in the sequel.

The partition function of system plus bath consists of a product of contributions
arising from the two normal modes, i.e. the center-of-mass and relative motion,
and thus reads
\begin{equation}
\label{eq:zsb0}
Z_\mathrm{SB} = \frac{L}{\hbar}\left[\frac{2\pi(m+M)}{\beta}\right]^{1/2}
                \frac{1}{2\sinh\left(\displaystyle\frac{\hbar\beta\bar\omega}{2}\right)}\,.
\end{equation}
From (\ref{eq:zb0}), (\ref{CB}) (\ref{eq:gabbrev})  and
(\ref{eq:zsb0}) it is straightforward to evaluate the
specific heat (\ref{eq:specificheat}) which becomes
\begin{equation}
\label{eq:cfree}
\frac{C}{k_\mathrm{B}} = \frac{1}{2}+g\left(\frac{\hbar\beta\bar\omega}{2}\right)
-g\left(\frac{\hbar\beta\omega}{2}\right)\,.
\end{equation}
The first term arises from a free particle while the second and third term describe the
change in specific heat due to the increase in the oscillator frequency from $\omega$
to $\bar\omega$ as given by (\ref{eq:omegabar}).

\begin{figure}
\begin{center}
\includegraphics[width=\columnwidth]{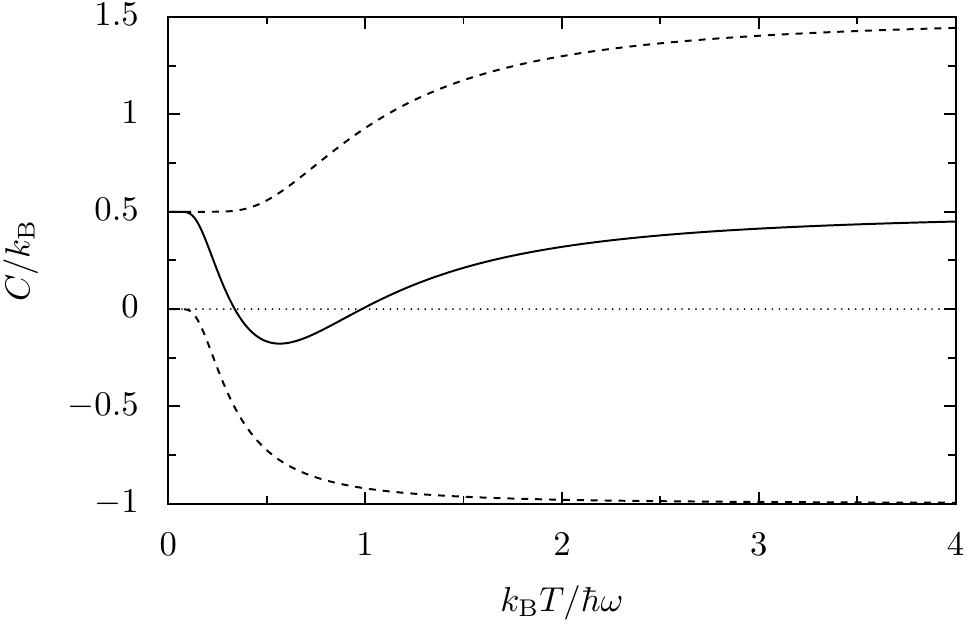}
\end{center}
\caption{The difference of specific heats as function of the temperature
for the case of a free particle coupled to a single oscillator bath
for the mass ratio $m/M=10$. The upper dashed curve corresponds to the
first two terms in (\ref{eq:cfree}), i.e. to the specific heat of the system shown
in Fig.~\ref{fig:toymodels}b while the lower dashed curve corresponds to the third
term in (\ref{eq:cfree}), i.e. to the negative of the specific heat of the bath
oscillator shown in Fig.~\ref{fig:toymodels}a.  The resulting difference is depicted
as solid line and displays a temperature region where it takes on negative values.}
\label{fig:freeparticle}
\end{figure}

In Fig.~\ref{fig:freeparticle} the contributions to (\ref{eq:cfree}) are
sketched. The upper dashed curve corresponds to the first two contributions
arising from system and bath. It contains the $k_\mathrm{B}/2$ from
the isolated free
particle and a contribution from the harmonic oscillator which is
strongly suppressed at low temperatures and reaches $k_\mathrm{B}$ for high
temperatures. The lower dashed curve corresponds to the third term in
(\ref{eq:cfree}). The main point to note is the relative shift in temperature
of the two contributions due to the change of the oscillator frequency. In the
presence of the system, the oscillator frequency is increased according to
(\ref{eq:omegabar}). As a consequence, there is a temperature window, where the
specific heat of system and bath is already significantly suppressed while this
is not yet the case for the bath oscillator alone.  In this regime, the
difference (\ref{eq:cfree}) of the specific heats can become negative.  Note
that in contrast to Fig.~\ref{fig:c} this temperature window does not extend
all the way down to zero temperature. This is explained by the fact that the
bath consists of only one oscillator so that the low-frequency oscillators
present in the Drude model (\ref{eq:drude}) on which Fig.~\ref{fig:c} is based
are missing.

\section{Harmonic oscillator}
\label{sec:ho}
After having explained the origin of a negative specific heat for a
free particle in contact with an environment we will now address the
question why such a scenario cannot be realized for a harmonic
oscillator. 
\subsection{Minimal environment}
To
this end, we determine the difference of specific heats for the
mechanical systems shown
in Figs.~\ref{fig:toymodels}a and \ref{fig:toymodels}c. The system frequency associated with the spring
constant $f_\text{S}$ is given by $\Omega=(f_\text{S}/M)^{1/2}$. The eigenfrequencies
of system plus bath are readily obtained as
\begin{equation}
\begin{aligned}
\omega_\pm^2 &= \frac{1}{2}\left(\frac{m+M}{M}\omega^2+\Omega^2\right)\\
&\qquad\pm
\left[\frac{1}{4}\left(\frac{m+M}{M}\omega^2+\Omega^2\right)^2-\omega^2\Omega^2\right]^{1/2}\,.
\end{aligned}
\end{equation}
One can show that $\omega_-\leq\omega\leq\omega_+$
where $\omega_{-}=\omega$ or $\omega_{+}=\omega$ for $m=0$ depending
on whether $\omega/\Omega >1$ or $\omega/\Omega <1$.
For increasing mass ratio $m/M$ and fixed frequency ratio
$\omega/\Omega$ both gaps from $\omega$ to $\omega_{-}$ and
$\omega_{+}$ widen.
In the limit $m\gg M$, at any fixed frequency ratio $\omega/\Omega$
one finds $\omega_-=(M/m)^{1/2}\Omega$ and $\omega_+=(m/M)^{1/2}\omega$.

\begin{figure}
\begin{center}
\includegraphics[width=\columnwidth]{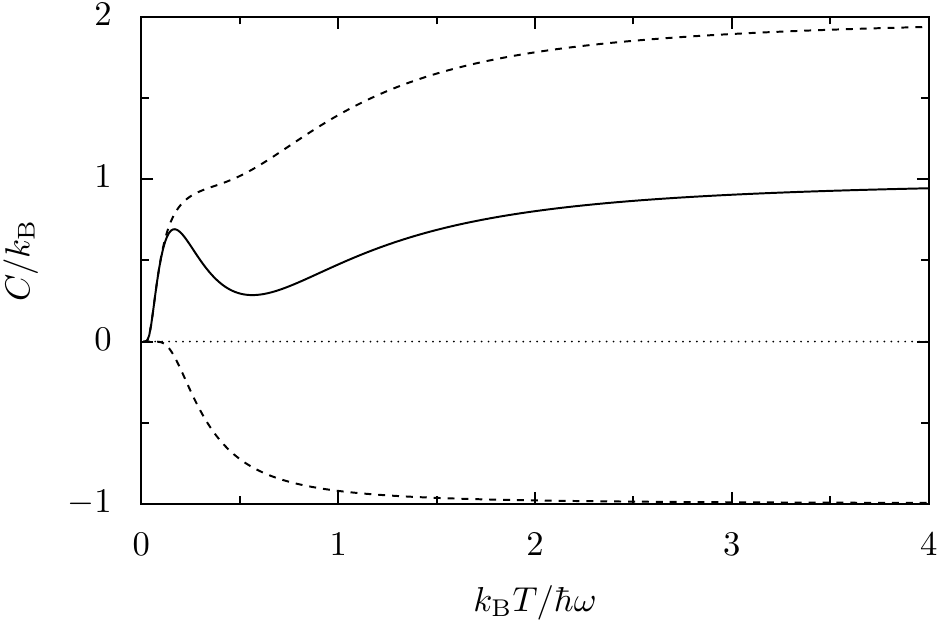}
\end{center}
\caption{The difference of specific heats as function of the temperature
for the case of a harmonic oscillator coupled to a single oscillator
bath. The upper dashed curve corresponds to the
first two terms in (\ref{eq:chos}), i.e. to the specific heat of the system shown
in Fig.~\ref{fig:toymodels}c while the lower dashed curve corresponds to the third
term in (\ref{eq:chos}), i.e. to the negative of the specific heat of the bath
oscillator shown in Fig.~\ref{fig:toymodels}a.  The resulting difference is depicted
as solid line and displays a dip for the chosen parameter values
$\Omega = \omega$ and $m/M=10$.}
\label{fig:oscillator}
\end{figure}

From (\ref{eq:partitionfunction}) and (\ref{eq:specificheat}) one then obtains
\begin{equation}
\label{eq:chos}
\frac{C}{k_\mathrm{B}} = g\left(\frac{\hbar\beta\omega_+}{2}\right)
+g\left(\frac{\hbar\beta\omega_-}{2}\right)
-g\left(\frac{\hbar\beta\omega}{2}\right)
\end{equation}
where $k_{\mathrm{B}}g$ is the specific heat of a harmonic oscillator
defined in Eqs. (\ref{CB}) and (\ref{eq:gabbrev}).
A typical scenario for the case of sufficiently well separated
frequencies is sketched in
Fig.~\ref{fig:oscillator}. Although for the harmonic oscillator a dip in the
specific heat may appear, no negative values can be obtained. The main difference
to the case of a free particle is the specific heat of the isolated system degree
of freedom: While for not too low temperatures the specific heat for the harmonic
oscillator equals $k_\mathrm{B}$ it is only half as large for the free particle.
The difference of the specific heats of the bath oscillator in the presence and
absence of the system degree of freedom may reach values up to $k_\mathrm{B}$,
thereby opening up the possibility of negative values of the specific heat for the
free particle but not for the harmonic oscillator.

\subsection{Drude bath}
The dip in the specific heat of the harmonic oscillator in contact
with the minimal environment can also be observed
for a bath giving rise to ohmic damping, i.e. where $\hat\gamma(0)>0$. For the ratio
of partition functions (\ref{eq:partitionfunction})  one finds
\begin{equation}
\label{eq:zho}
Z = \frac{1}{\hbar\beta\Omega}\prod_{n=1}^\infty \frac{\nu_n^2}
{\nu_n^2+\nu_n\hat\gamma(\nu_n)+\Omega^2}\,,
\end{equation}
where $\nu_n=2\pi n/\hbar\beta$ are the Matsubara frequencies.

\begin{figure}
\begin{center}
\includegraphics[width=\columnwidth]{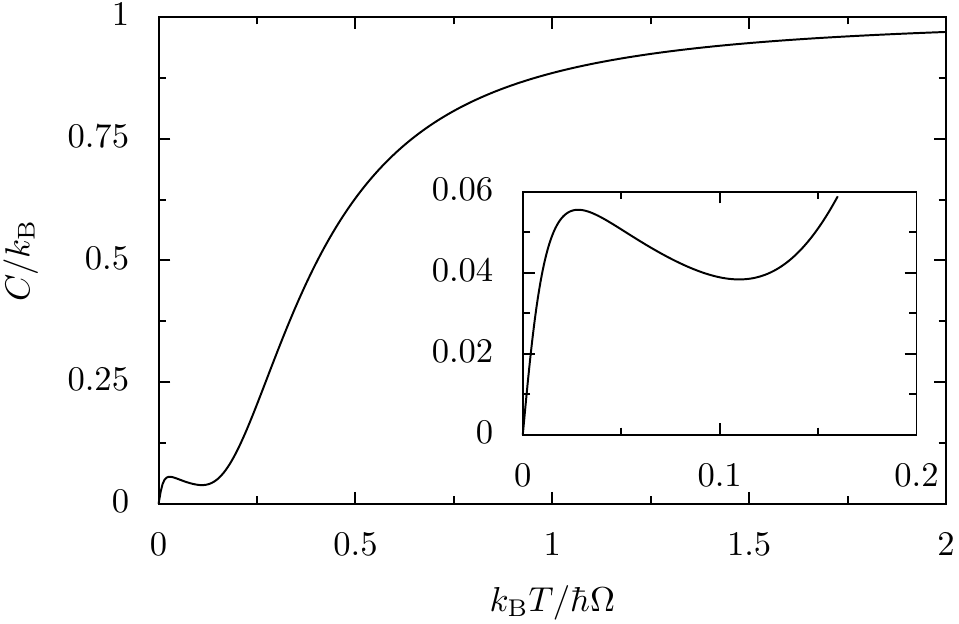}
\end{center}
\caption{Specific heat (\ref{eq:cho}) of a damped harmonic oscillator for
Drude damping with $\gamma=5\Omega$ and $\omega_\mathrm{D}=0.1\Omega$. The
inset shows a close-up of the low-temperature regime.}
\label{fig:cho}
\end{figure}

We specifically consider a Drude model characterized by the damping kernel
(\ref{eq:drude}) with a cutoff frequency $\omega_\mathrm{D}$. Introducing the
quantities
\begin{equation}
\Lambda_\mathrm{D}=\frac{\hbar\beta\omega_\mathrm{D}}{2\pi}
\end{equation}
and
\begin{equation}
\Lambda_i=\frac{\hbar\beta\lambda_i}{2\pi},\qquad i=1,2,3
\end{equation}
with
\begin{equation}
\label{eq:cubic}
\begin{split}
x^3+\omega_\mathrm{D}x^2+(\gamma\omega_\mathrm{D}+\Omega^2)x
+\omega_\mathrm{D}\Omega^2\\
= (x-\lambda_1)(x-\lambda_2)(x-\lambda_3)\,,
\end{split}
\end{equation}
we obtain from (\ref{eq:specificheat}) with (\ref{eq:zho})
\begin{equation}
\begin{aligned}
\label{eq:cho}
\frac{C}{k_\mathrm{B}} &= 1
+\Lambda_1^2\psi'(1-\Lambda_1)
+\Lambda_2^2\psi'(1-\Lambda_2)\\
&\qquad+\Lambda_3^2\psi'(1-\Lambda_3)
-\Lambda_\mathrm{D}^2\psi'(1-\Lambda_\mathrm{D})\,.
\end{aligned}
\end{equation}
Here, $\psi'(x)$ denotes the trigamma function. The specific heat (\ref{eq:cho})
is shown in Fig.~\ref{fig:cho} for $\gamma=5\Omega$ and $\omega_\mathrm{D}=0.1\Omega$
where $\Omega$ is the frequency of the system oscillator. In contrast to the
undamped case, the specific heat increases linearly with temperature with a slope
proportional to the damping constant. Then, for sufficiently small
cutoff frequency  the specific heat goes through
a dip at low temperatures
before it asymptotically approaches its
high-temperature value $k_\mathrm{B}$.

\section{Density of states}
\begin{figure}
\begin{center}
\includegraphics[width=\columnwidth]{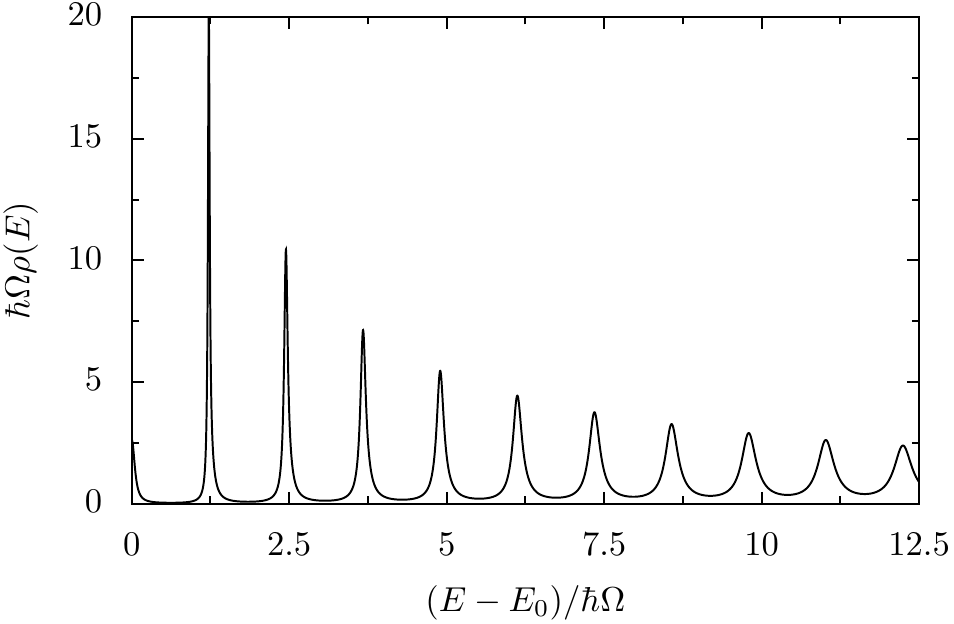}
\end{center}
\caption{Density of states (\ref{eq:rho}) of a damped harmonic oscillator for
Drude damping with $\gamma=5\Omega$ and $\omega_\mathrm{D}=0.1\Omega$. For these
parameters, (\ref{eq:cubic}) yields a resonance frequency of about $1.224\Omega$.
A delta function contribution of weight one present at the ground state energy
$E_0$ of the damped harmonic oscillator is not shown \cite{footnote2}.}
\label{fig:zddrude_fft}
\end{figure}

One can use the partition function (\ref{eq:partitionfunction}) to formally
define an effective density of states $\rho(E)$ of the system by means of the
relation \cite{hanke95}
\begin{equation}
\label{eq:rho}
Z = \int_0^\infty\mathrm{d}E \rho(E)\exp(-\beta E)\,.
\end{equation}
For the free particle, it was found in Ref.~\cite{hangg08} that the density of
states obtained from (\ref{eq:rho}) can become negative for appropriately
chosen environments, e.g.  a Drude model with $\omega_\mathrm{D}<\gamma$. For
an environment consisting of a single bath mode like in the minimal models
discussed in Sect.~\ref{sec:free} and \ref{sec:ho} it was demonstrated, that
the appearance of negative contributions to the effective density of states is
generic. In particular, the minimal model for a harmonic oscillator coupled to
an environmental mode displayed in Fig.~\ref{fig:toymodels}c leads to negative
delta functions in the density of states, see Ref. \cite{hangg08}.
This is in contrast to our finding
for the specific heat which remains always positive for the harmonic
oscillator.  Interestingly, for a heat bath exhibiting a continuous
distribution of bath oscillators, one obtains again a positive density of
states despite of the fact that dips in the specific heat like the one shown in
Fig.~\ref{fig:cho} can be observed.  Fig.~\ref{fig:zddrude_fft} displays the
density of states for the same parameters as employed in Fig.~\ref{fig:cho}.

An important difference between the specific heat and the density of states as
defined by (\ref{eq:rho}) lies in the fact that the latter cannot be
interpreted in terms of a difference of two densities of states. We recall that
such an interpretation was possible for the specific heat only because
according to (\ref{eq:specificheat}) it depends linearly on the logarithm of
the partition function (\ref{eq:partitionfunction}).  The specific heat shares
this property with other thermodynamic quantities like the internal energy and
the free energy. The absence of a logarithm of the partition function in
(\ref{eq:rho}) indicates that the effective density of states does not lend
itself to an interpretation in terms of the difference of two densities of
states. Despite the fact that the effective density of states of a damped
harmonic oscillator in the weak-coupling limit, i.e. when the damping strength
represents the smallest frequency scale, displays resonances at the expected
energies and even yields the correct level widths \cite{ingol02}, it therefore
remains unclear whether the meaning of the effective density of states goes
beyond that of a merely formal notion.

\section{Conclusions}
For a free damped particle, the specific heat based on the effective partition
function (\ref{eq:partitionfunction}) can become negative. We have demonstrated that
this surprising behavior does not endanger the thermodynamic stability of the
damped system. Instead, the specific heat should be interpreted as the change
of the specific heat of the environment when a system degree of freedom is
attached to it. For a damped harmonic quantum oscillator, the difference of specific
heats cannot become negative but may display a dip instead. The
difference in the
behavior of the free particle and the harmonic oscillator can be traced back to
the specific heat of the uncoupled system which for the free particle is
smaller by a factor of two. All those quantities that are obtained from the
logarithm of the partition function by means of a linear operation can be
interpreted as differences between the corresponding quantities of the total system
plus bath and of the bath alone.  This reasoning though does not apply to the
density of states which is the inverse Laplace transform of the partition
function itself, see (\ref{eq:rho}). By this transformation the ratio of two
partition functions yields a complex quantity that obviously cannot be
interpreted as a difference of the densities of states of the total system and
of the environment.

\begin{acknowledgments}
One of us (GLI) is indebted to Hermann Grabert for a discussion
which motivated the present work. The manuscript was completed
during a stay of GLI at the Laboratoire Kastler Brossel with financial support
by the European Science Foundation (ESF) within the activity `New
Trends and Applications of the Casimir Effect' (\url{www.casimir-network.com}).
This work was supported by DFG via research center SFB-486, project
A10 and the German Excellence Initiative via the {\it Nanosystems
Initiative Munich} (NIM).
\end{acknowledgments}


\begin{thebibliography}{00}
\bibitem{footnote1}
In Ref.~\cite{hangg08} this quantity was referred to as $C^Z$.
\end{thebibliography}
\end{document}